\documentclass[twocolumn, prl, superscriptaddress]{revtex4}
\usepackage{amsmath,amssymb}
\usepackage{textcomp}
\usepackage{graphicx,color}
\usepackage{ulem}
\newcommand{\hc}{\color{blue}}

\begin{document}
\title{Magnetic field dependence of Pauli spin blockade: a window into the sources of spin relaxation in silicon quantum dots}

\author{G. Yamahata}
\affiliation{Quantum Nanoelectronics Research Center, Tokyo Institute of Technology, 2-12-1 O-okayama, Meguro-ku, Tokyo 152-8552, Japan}
\affiliation{Department of Physics, Harvard University, Cambridge, Massachusetts 02138, USA}

\author{T. Kodera}
\affiliation{Quantum Nanoelectronics Research Center, Tokyo Institute of Technology, 2-12-1 O-okayama, Meguro-ku, Tokyo 152-8552, Japan}
\affiliation{Institute for Nano Quantum Information Electronics, The University of Tokyo, Tokyo 153-8505, Japan}
\affiliation{PRESTO, Japan Science and Technology Agency (JST), 4-1-8 Honcho Kawaguchi, Saitama, Japan}

\author{H. O. H. Churchill}
\affiliation{Department of Physics, Harvard University, Cambridge, Massachusetts 02138, USA}

\author{K. Uchida}
\affiliation{Department of Physical Electronics, Tokyo Institute of Technology, 2-12-1 O-okayama, Meguro-ku, Tokyo 152-8552, Japan}

\author{C. M. Marcus}
\affiliation{Department of Physics, Harvard University, Cambridge, Massachusetts 02138, USA}

\author{S. Oda}
\affiliation{Quantum Nanoelectronics Research Center, Tokyo Institute of Technology, 2-12-1 O-okayama, Meguro-ku, Tokyo 152-8552, Japan}

\date{\today}

\begin{abstract}
We investigate spin relaxation in a silicon double quantum dot via leakage current through Pauli blockade as a function of interdot detuning and magnetic field. A dip in leakage current as a function of magnetic field on a $\sim 40$ mT field scale is attributed to spin-orbit mediated spin relaxation. On a larger ($\sim 400$ mT) field scale, a peak in leakage current is seen in some, but not all, Pauli-blocked transitions, and is attributed to spin-flip cotunneling. Both dip and peak structure show good agreement between theory and experiment. 
\end{abstract}

\maketitle
Electron spins confined in semiconductor quantum dots (QDs) are attractive candidates for quantum information processing \cite{ldqubit}. 
Coherent manipulation of individual and coupled electron spin states has been mainly investigated in GaAs-based double QD (DQD) devices \cite{kop1,pet1,slBqubit}. However, nuclear spins of the host material cause decoherence of the electron spin via strong hyperfine coupling \cite{kae1}. To reduce this effect, group IV materials, such as carbon, silicon (Si), and silicon-germanium (SiGe), have been investigated \cite{chucnt1,nswdqd,HWJ1,coreshell,eriT1} because their most abundant isotopes have zero nuclear spin. Silicon systems, in particular, have an advantage for future integration because of their compatibility with conventional Si metal-oxide-semiconductor devices.

\begin{figure}
\begin{center}
\includegraphics[width= 3 in]{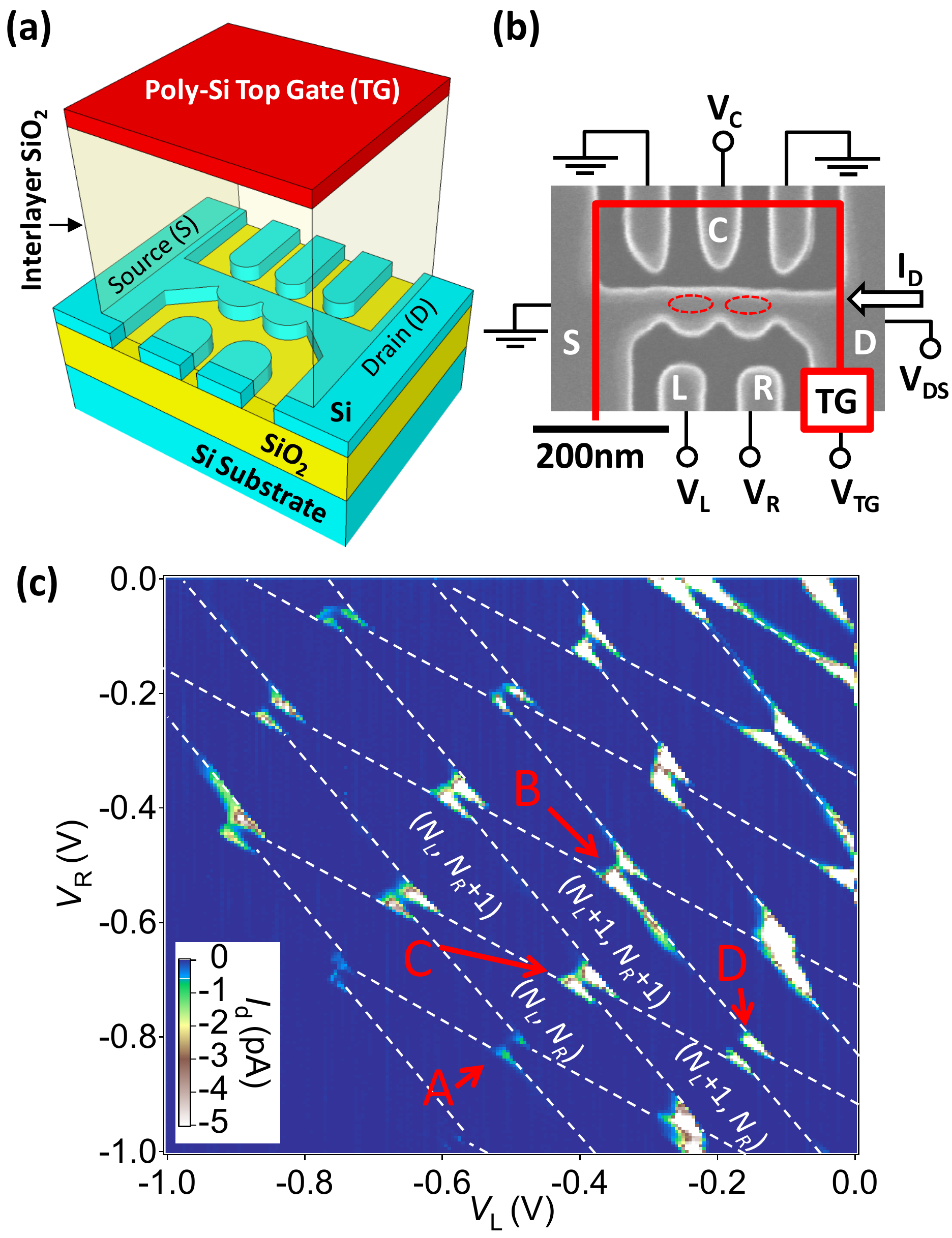}
\end{center}
\caption{(color online) (a) Schematic of the silicon double quantum dot (Si DQD). (b) Scanning electron microscope image of the Si DQD before the top gate formation. The two side gates located next to side gate C are grounded. (c) Charge stability diagram of the Si DQD as a function of $V_{\mathrm{L}}$ and $V_{\mathrm{R}}$ at zero magnetic field, where $V_{\mathrm{ds}} = - 2$ mV, $V_{\mathrm{TG}} = 0.90$ V, and $V_{\mathrm{C}} = -1.72$ V. The white dotted lines are boundaries of the stable charge states. The charge numbers in the left and right QDs are $N_{\mathrm{L}}$ and $N_{\mathrm{R}}$, respectively.}
\label{f1}
\end{figure}

Toward spin qubits in Si systems, it is necessary to understand the spin relaxation mechanism. Pauli spin blockade (PSB) \cite{onospin, jspin} is a valuable tool for investigating spin relaxation in confined systems. In DQDs of several materials, the spin relaxation mechanism has been characterized by analyzing the leakage current in the PSB regime \cite{kop2, InAs1, chucnt2, kodera1}, where hyperfine interaction and/or spin-orbit interaction dominate the spin relaxation. For Si systems, a PSB has been reported for a DQD in metal-semiconductor-oxide structures and an electrostatically formed DQD in Si/SiGe heterostructures \cite{erispin, nttspin}. However, the relaxation mechanism in Si DQDs has not yet been experimentally clarified. More recently, magnetic field dependences of the leakage current in a PSB regime have been demonstrated in a pure Si DQD \cite{NSWsb1}, where a current peak was explained by field-dependent cotunneling.

In this Letter, we investigate leakage current in a PSB regime using a lithographically defined Si DQD. By changing magnetic field, we observed a dip of the leakage current at zero magnetic field, presumably the result of spin-orbit-mediated spin relaxation. In addition, magnetic field dependences at a different charge triple point exhibit a leakage current peak at zero magnetic field. This peak can be understood as a signature of spin-flip cotunneling processes.

Figure \ref{f1}(a) shows a schematic of a Si DQD. Three constrictions between the source (S) and drain (D), and five side gates were patterned by electron beam lithography on a 60-nm-thick (100) Si-on-insulator (SOI) layer, where the thickness of the buried oxide was 400 nm. Reactive ion etching was used to transfer the resist pattern onto the SOI, followed by formation of the gate oxide via thermal oxidation for 30 min at 1000~\textcelsius \ and low-pressure chemical vapor deposition (LPCVD). Then, a wide poly-Si top gate (TG) formed by LPCVD was used as an ion implantation mask for the formation of the n-type S and D regions. Finally, 300-nm-thick aluminum contact pads were formed by electron beam evaporation. Figure \ref{f1}(b) shows a scanning electron microscope image of the device, where the DQD is defined by tunnel barriers at the three constricted regions \cite{me2}. 

Electrons were attracted to the Si (100) surface by applying a positive TG voltage, $V_{\mathrm{TG}}$. Electrochemical potentials of the left and right QDs  were modulated by applying voltages $V_{\mathrm{L}}$ and $V_{\mathrm{R}}$ to side gates L and R. The tunnel coupling between the two QDs was controlled by voltage $V_{\mathrm{C}}$ applied to side gate C. All measurements were carried out in a $^{3}$He refrigerator with a base temperature of 250 mK.

The honeycomb charge stability [Fig.~\ref{f1}(c)] reflects the formation of a DQD \cite{dqd1}. Charging energies of the left and right QDs were estimated to be $10.7$ and $11.0$ meV, respectively, from the spacings of the Coulomb peaks, implying that the QDs have almost the same size. In addition, from the distribution of the current peaks due to resonant tunneling at triple point A in Fig.~\ref{f1}(c), the quantum level spacing, $\Delta E$, of the left and right QD was estimated to be 310 and 260 $\mu$eV, respectively \cite{EPAPS}. In confirmation, $\Delta E$ can be approximated as $\Delta E = {h^2}/{8\pi m^{*}A}$,
where $m^{*}$ gives effective mass here, $h$ is Planck's constant, and $A$ is the area of the QD \cite{kouqd1}, with spin and valley degeneracies included. This equation determines $\Delta E$ to be between 260 and 380 $\mu$eV for our device geometry \cite{EPAPS}, in good agreement with the experimental estimation. We conclude that the QD is formed between the two constricted regions indicated by the ovals in Fig.~\ref{f1}(b).

\begin{figure}
\begin{center}
\includegraphics[width=250pt, clip]{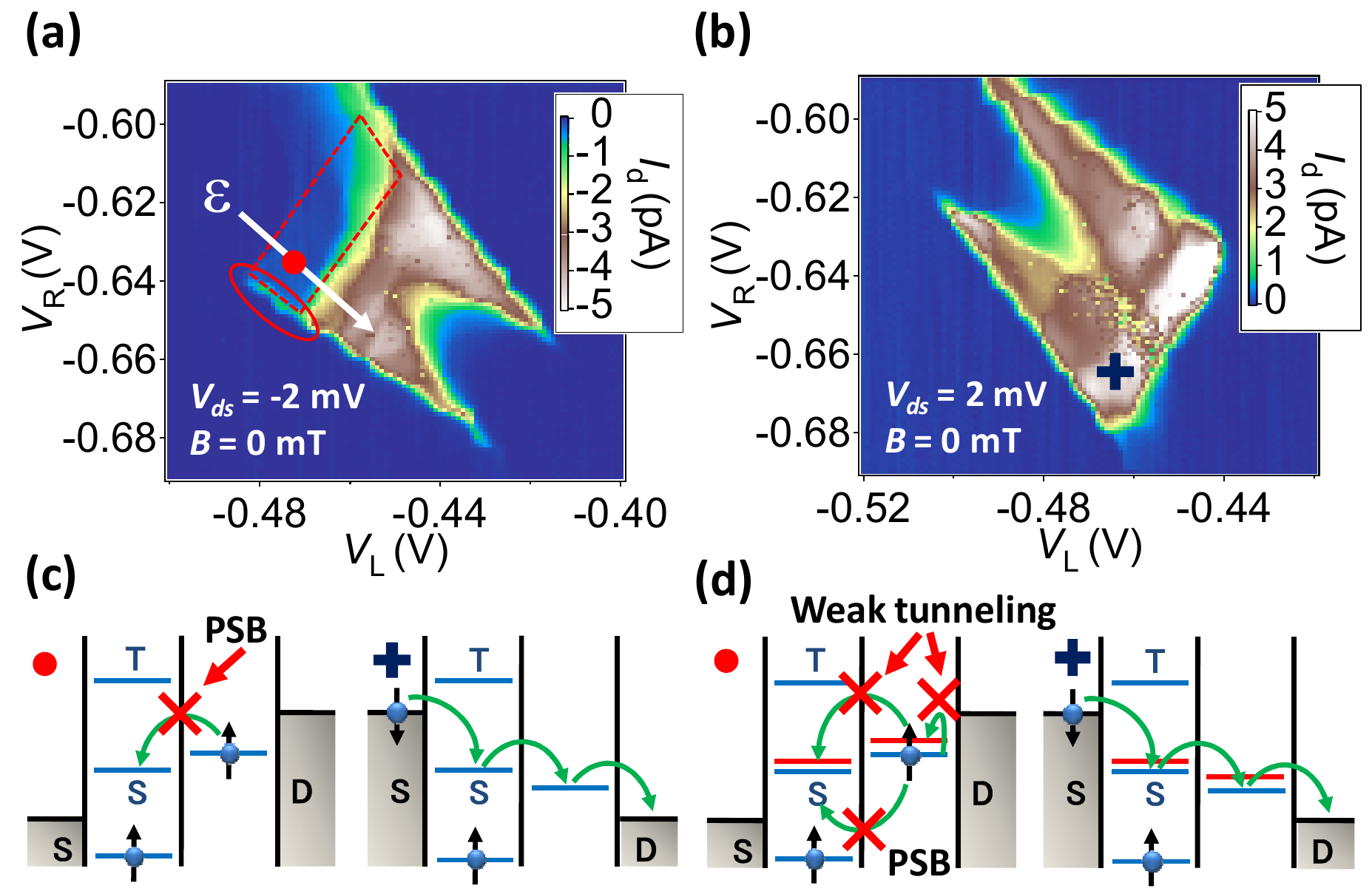}
\end{center}
\caption{(color online) (a) Triple point B shown in Fig.~\ref{f1}(c) with negative bias, where $V_{\mathrm{ds}} = -2$ mV, $V_{\mathrm{TG}} = 0.97$ V, and $V_{\mathrm{C}} = -1.76$ V. The PSB appears only for this polarity. Here $\epsilon $ is the detuning axis. (b) The same triple point as in {\hc (a)} under a positive bias ($V_{\mathrm{ds}} = 2 $ mV). (c) Energy diagrams of a Si DQD at the circle marked in a (the left diagram) and at the blue cross marked in b (the right diagram), where the valley degeneracy is assumed to be lifted. (d) The same diagram as {\hc (c)} without an assumption that lifting of the valley degeneracy is small. Intra-dot and inter-dot tunnelings between different valleys are assumed to be weak so that the PSB is not lifted.}
\label{f2}
\end{figure}

Current rectification in DQDs due to a PSB appears at a triple point with only one bias polarity \cite{jspin}. We observed such current rectification with a negative bias voltage at triple point B in Fig.~\ref{f1}(c), as indicated by the trapezoid in Fig.~\ref{f2}(a), whereas no current rectification appeared with positive bias as shown in Fig.~\ref{f2}(b). In addition, the current rectification is lifted along the outer edge of the PSB regime indicated by the circle in Fig.~\ref{f2}(a) because of electron exchanges between the DQD and the right lead, comparable to PSB seen in GaAs DQDs \cite{jspin}.

Since Si DQDs normally have doubly degenerate valleys due to confinement in the direction perpendicular to the Si surface, the valley degeneracy could lift a PSB. However, the fact that a PSB is observed indicates either a lifting of valley degeneracy or weak tunneling between valleys \cite{das1}. In the former case, once two spins occupy the (1, 1) triplet state as shown in Fig.~\ref{f2}(c), the current flow is suppressed due to the PSB until relaxation from (1, 1) triplet to (1, 1) singlet occurs. In the latter case, even if degenerate valleys exist as shown in Fig.~\ref{f2}(d), the PSB is not lifted because intra-dot and inter-dot tunnelings between valleys are weak.

PSB features were observed at adjacent triple points, marked B, C, and D in Fig.~\ref{f1}(c).  This is not expected for simple spin-$\frac{1}{2}$ PSB. Since the DQD has many electrons, spin-$\frac{3}{2}$ ground states can exist, leading to scenarios for consecutive PSB \cite{jspin}.  Blockade where valley degeneracy plays a role can also lead to consecutive PSB-like features. Even when a spin doublet is formed in DQDs, the current flow could be suppressed  because of weak tunneling between valleys discussed above \cite{EPAPS}. 

\begin{figure}
\begin{center}
\includegraphics[width=255pt, clip]{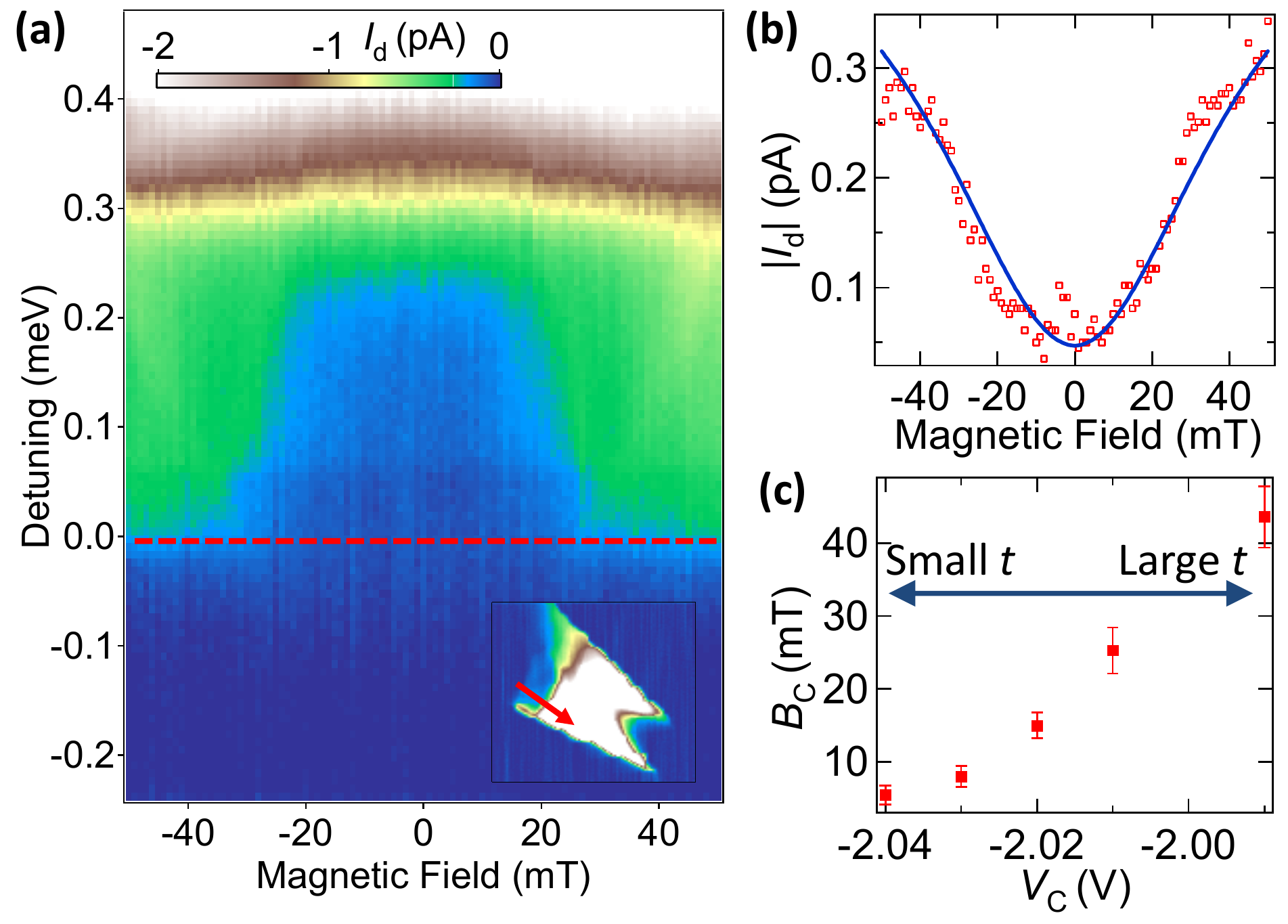}
\end{center}
\caption{(color online) (a) Leakage current in the PSB regime as a function of magnetic field applied perpendicularly to the DQD and detuning, where $V_{\mathrm{ds}} = -2$ mV, $V_{\mathrm{TG}} = 0.97$ V, and $V_{\mathrm{C}} = -1.99$ V. Inset: Magnified plot of triple point C in Fig.~\ref{f1}(c), where the arrow corresponds to the detuning axis in the main figure. (b) Current along the dashed line in {\hc (a)} denoted by the squares, and the fit to the data indicated by the blue line. (c) Values of $B_{\mathrm{C}}$ extracted from the fit as a function of $V_{\mathrm{C}}$. Large $V_{\mathrm{C}}$ corresponds to a large inter-dot tunnel coupling {\hc $t$.}}
\label{f3}
\end{figure}

Figure \ref{f3}(a) shows the leakage current in the PSB regime at triple point C in Fig.~\ref{f1}(c) as a function of magnetic field $B$ applied normally to the DQD with a detuning, $\epsilon $, corresponding to the arrow shown in the inset. A strong current dip was observed at $B = 0$, whereas the current with opposite bias does not change as a function of magnetic fields \cite{EPAPS}. Similar current dips have been observed for DQDs in InAs nanowires \cite{InAs1, Nadj1} and carbon nanotubes \cite{chucnt2} and can be attributed to spin-orbit induced relaxation \cite{SSO1}, which is suppressed at $B = 0$ due to a Van Vleck cancellation \cite{InAs1, kae2}.  A Lorentzian line shape, $I_{\mathrm{fit}} = I_{\mathrm{max}}\{1-{8B_{\mathrm{C}}^{2}}/{9(B^{2} + B_{\mathrm{C}}^{2})}\}$ with characteristic width $B_{\mathrm{C}}$,  is predicted theoretically \cite{SSO1}. The squares in Fig.~\ref{f3}(b) correspond to the absolute values of the leakage current in the PSB regime along the dashed line in Fig.~\ref{f3}(a). Fits to the Lorentzian form (the blue curve in Fig.~\ref{f3}(b)) yield good agreement between theory and experiment. Furthermore, as the inter-dot tunneling between the two QDs is enhanced by changing $V_{\mathrm{C}}$, the value of $B_{\mathrm{C}}$ extracted from the fit increases, as plotted in Fig.~\ref{f3}(c). This result is also consistent with the theory, which predicts $B_{\mathrm{C}}$ proportional to inter-dot tunnel coupling \cite{SSO1}. These results suggest that spin-orbit effects dominate spin relaxation in these devices. 

Another possible mechanism leading to a dip in current leakage around $B=0$ is  spin-valley blockade with short-range disorder \cite{palyi2}, where the current dip as a function of magnetic-field-induced valley splitting is predicted. However, we have no independent evidence that the required B-dependent valley splitting exists. The physics of the valley in Si DQDs deserves further experimental and theoretical study.

\begin{figure}
\begin{center}
\includegraphics[width=250pt]{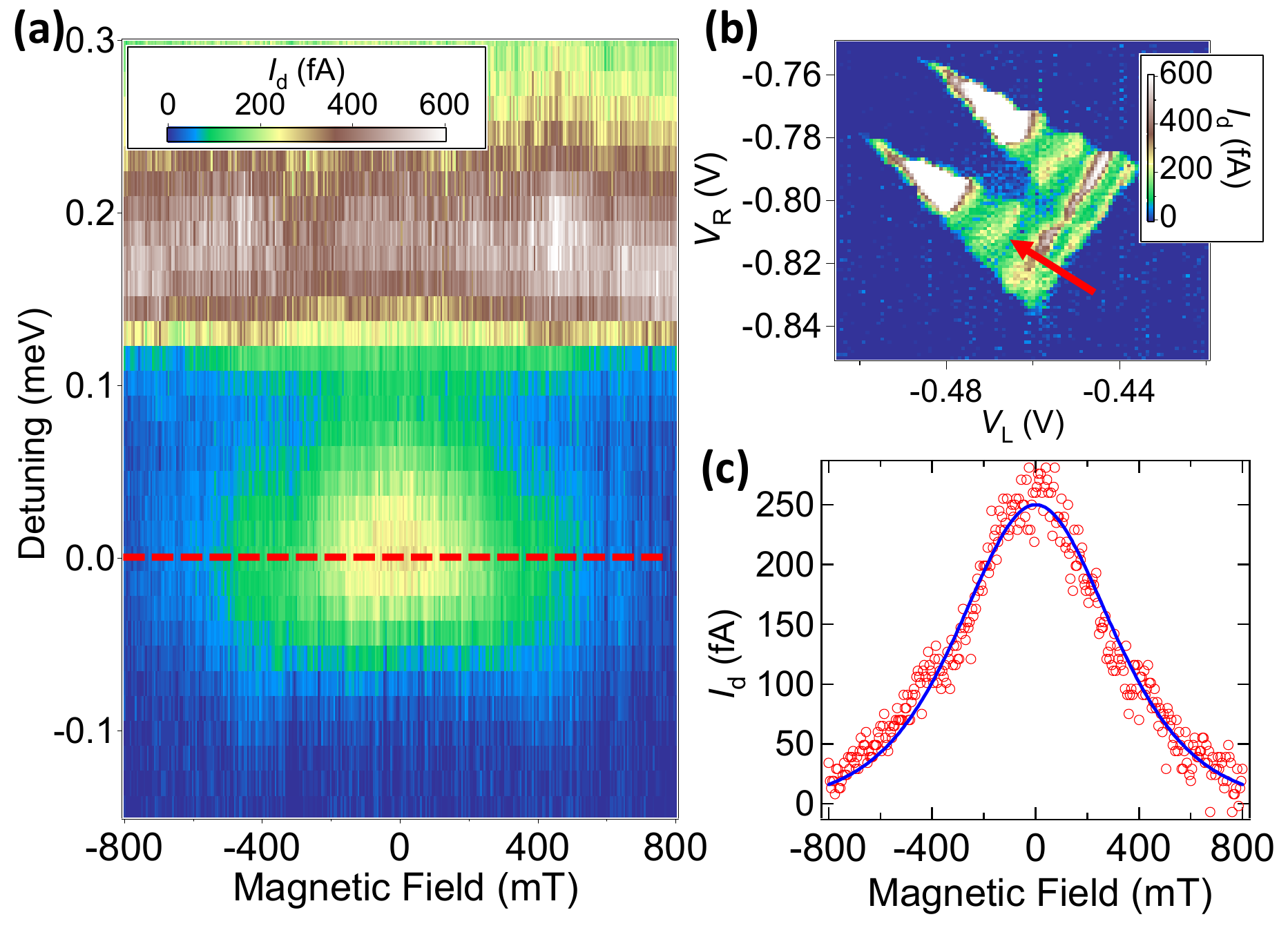}
\end{center}
\caption{(color online) (a) Leakage current in the PSB regime as a function of magnetic field applied perpendicularly to the DQD and the detuning, where$V_{\mathrm{ds}}= 2$ mV, $V_{\mathrm{TG}} = 0.968$ V, and $V_{\mathrm{C}} = -1.925$ V.  (b) Magnified plot of triple point A in Fig.~\ref{f1}(c), where the arrow corresponds to the detuning axis in (a). (c) Current along the dashed line in (a) denoted by the circles, and the fit to the data indicated by the blue line. }

\label{f4}
\end{figure}

For some triple points, we observe a peak, rather than dip, in PSB leakage current on a larger field scale. As an example, the field dependence of the leakage current at triple point A in Fig.~\ref{f1}(c) is shown in Fig.~\ref{f4}(a). The arrow in the magnified plot of triple point A shown in Fig.~\ref{f4}(b) corresponds to the detuning axis in Fig.~\ref{f4}(a). Among the 15 triple points that show PSB [Fig.~\ref{f1}(c)], nine show a zero-field current dip and two show a peak. We also observed current peaks outside a current dip in some cases. 

In GaAs DQDs, zero-field peaks in leakage current were attributed to hyperfine-induced spin relaxation \cite{kop2, naz1}. However, the contribution of the hyperfine interaction should be small in Si systems, because the dominant $^{28}$Si atoms have zero nuclear spin. 
Using 4.7\,\% natural abundance of $^{29}$Si and lithographic device dimensions \cite{EPAPS} gives an expected number $N$ of nuclear spins in a Si DQD to be $2$\,-\,$3 \times 10^{4}$, corresponding to a fluctuating Overhauser field magnitude $B_{\mathrm{nuc}} = |A|/g\mu _{\mathrm{B}}\sqrt{N} \sim 10$\,-\,$15\, \mu$T, where the hyperfine coupling constant $|A| \sim 0.2\, \mu$eV from NMR measurements \cite{sinmr} and $g\sim2$ for electrons in Si. Since the peak width in Fig.~\ref{f4}(c) is larger than $B_{\mathrm{nuc}}$  by a factor of 10$^4$, the mechanism of the current peaks at $B = 0$ is not explained by hyperfine interaction.

Similar peaks were also seen in Si DQD in Ref.~\cite{NSWsb1}, where the peak is well described by spin-flip cotunneling \cite{coish1}. When $k_{\mathrm{B}}T>t$ ($k_{\mathrm{B}}$ is Boltzmann's constant and $t$ is the inter-dot tunnel coupling), the spin-flip cotunneling current is given by $I_{\mathrm{cot}} = 4ecg\mu _{\mathrm{B}}B/3{\mathrm{sinh}}(g\mu _{\mathrm{B}}B/k_{\mathrm{B}}T)$ with $c = h[(\Gamma _{\mathrm{R}}/(\Delta -\epsilon ))^2 + (\Gamma _{\mathrm{L}}/(\Delta + \epsilon -2U' - 2eV_{\mathrm{ds}}))^2]/\pi$ where $\Gamma _{\mathrm{L(R)}}$ is the coupling of the lead to the left (right) dot, $\Delta$ is the depth of the two-electron level \cite{qassemi1}, and $U'$ is inter-dot charging energy. Since we observed clear resonant tunneling peaks, $\Gamma _{\mathrm{L(R)}}$ is larger than $t$ \cite {fuji2}. In addition, if $\Gamma _{\mathrm{L(R)}}>t>k_{B}T \sim 21~\mathrm{\mu eV}$, the current would be much larger than the observed current shown in Fig.~\ref{f4}(b). As a result, $k_{\mathrm{B}}T > t$ so that $I_{\mathrm{cot}}$ can be used to fit the current peak. The blue curve in Fig.~\ref{f4}(c) is $I_{\mathrm{cot}}$, which has a good agreement with the data by using $T \sim 250~\mathrm{mK}$, yielding $g \sim 2.3$ and $c \sim 54~\mathrm{kHz/\mu eV}$. Since the current does not vary much along the base of the triangle in Fig.~\ref{f4}(b), we assume $\Gamma _{\mathrm{L}} \sim \Gamma _{\mathrm{R}} \equiv \Gamma$. By using expression of $c$ with $\Delta \sim 1~ \mathrm{meV}$, $\epsilon \sim 0~\mathrm{meV}$, $U' \sim 1~\mathrm{meV}$, and $eV_{\mathrm{ds}} \sim 2~\mathrm{meV}$ estimated from the bias triangle shown in Fig.~\ref{f4}(b), we extracted $\Gamma \sim 26~\mathrm{\mu}$eV. Furthermore, $t$ can be extracted  to be about $0.3~\mu$eV from the unblocked resonant tunneling peak current ($\sim 0.6~\mathrm{pA}$) with Eq.~(15) in Ref.~\cite{dqd1}. These values are similar with those in Ref.~\cite{NSWsb1} and in an experimentally reasonable range so that the spin-flip cotunneling processes are most likely the mechanism of the peak. It should be noted that, as for the dip in Fig.~3, spin-valley blockade with disorder could also explain the peak, but again we have at present no evidence of the required field-dependent valley splitting \cite{pburkard}.

\begin{acknowledgments}
 GY and TK contributed equally to this work. We thank W. A. Coish, G. Burkard, A. P{\'{a}}lyi, C. Barthel, J. Medford, and F. Kuemmeth for valuable discussions and K. Usami, T. Kambara, R. Suzuki, and T. Hiramoto for device fabrication. This work was partly supported by a Grant-in-Aid for Scientific Research from the Ministry of Education, Culture, Sports, Science, and Technology of Japan (No. 21710137, No. 19206035, and No. 22246040), JST-PRESTO, and Special Coordination Funds for Promoting Science and Technology.
\end{acknowledgments}

\end{document}